\documentclass[pra,showpacs,twocolumn,nofootinbib]{revtex4}
\usepackage{graphicx}
\usepackage{bm,amsmath}
\usepackage{dcolumn}
\usepackage
{hyperref}

\newcommand{\sref}[1]{Sec. \ref{#1}}
\newcommand{\Eref}[1]{Eq.~(\ref{#1})}
\newcommand{\tref}[1]{Table~\ref{#1}}

\begin{document}

\title{Sensitivity coefficients to $\alpha$-variation for fine-structure transitions
in Carbon-like ions}
\author{M. G. Kozlov}
\affiliation{Petersburg Nuclear Physics Institute, Gatchina 188300,
             Russia}
\author{I. I. Tupitsyn}
\affiliation{St.~Petersburg State University, Petrodvorets, Russia}
\author{D. Reimers}
 \affiliation{Hamburger Sternwarte, Universit\"{a}t Hamburg, Gojenbergsweg 112,
              D-21029 Hamburg, Germany}
\date{
\today}


\begin{abstract}
We calculate sensitivity coefficients to $\alpha$-variation for the
fine-structure transitions (1,0) and (2,1) within $^3P_J[2s^2 2p^2]$
multiplet of the Carbon-like ions C~{\sc i}, N~{\sc ii}, O~{\sc
iii}, Na~{\sc vi}, Mg~{\sc vii}, and Si~{\sc ix}. These transitions
lie in the far infrared region and are in principle observable in
astrophysics for high redshifts $z\sim 10$. This makes them very
promising candidates for the search for possible $\alpha$-variation
on a cosmological timescale. In such studies one of the most
dangerous sources of systematic errors is associated with isotope
shifts. We calculate isotope shifts with the help of relativistic
mass shift operator and show that it may be significant for C~{\sc
i}, but rapidly decreases along the isoelectronic sequence and
becomes very small for Mg~{\sc vii} and Si~{\sc ix}.
\end{abstract}

\maketitle

\subsection{Introduction} \label{sec_intro}

Some of the extensions of the Standard Model predict small
space-time variations of such fundamental constants as the
fine-structure constant $\alpha=e^2/(\hbar c)$ and
electron-to-proton mass ratio $\mu=m_e/m_p$. These variations are
now intensively searched for both in astrophysical data and in
laboratory experiments. For this purpose two, or more lines with
different dependencies on the fundamental constants are compared to
each other at different times. Laboratory experiments provide
extremely high sensitivity to frequency shifts, while the typical
timescale is of the order of one year. The astrophysical
extragalactic observations allow studying time intervals comparable
to the lifetime of the universe ($\sim 10^{10}$ years), but with
much lower accuracy. Different theoretical models predict different
behavior of the constants from monotonic in time, to oscillations
and sharp changes. The latter could take place, for example, during
the transition from matter dominated to dark energy dominated
universe.


We can conclude that astrophysical and laboratory methods are
complementary and equally important. Recent progress and
perspectives of the laboratory searches were summarized in
\cite{FK07c}. At present all laboratory experiments give only strong
upper bounds on the time-variation of fundamental constants.
Situation in astrophysics is less clear. Some of the observations
indicate non-zero variation at a few-sigma level
\cite{MWF03,RBH06,LML07,LMK08}, while other results are consistent
with zero variation \cite{QRL04,CSP06,MFMH08}. For more references
and the discussion of the most recent developments see online
material from the Workshop \cite{Perim08}.

Controversial astrophysical results may indicate some systematic
effects. One of the possible systematic frequency shifts can arise
from the different velocity distributions of different species in
molecular clouds, i.e. the so-called Doppler noise
\cite{KCL05,LRK08}. Another potentially dangerous systematic effect
may be associated with the cosmological evolution of the isotope
abundances, which can lead to the time-dependent isotope shifts of
atomic transitions \cite{Lev94,MWF03}. The former effect can be
suppressed when different lines of the same specie are used to study
possible variation of constants. The latter systematics is absent in
microwave spectra of molecules, where lines of different isotopes
are well resolved. Alternatively, one can use either atoms with
single stable isotope, or specific combinations of atomic
frequencies, which are not sensitive to isotope shifts
\cite{KKBD04}. In order to apply this latter method one needs to
know accurate values of the isotope shift coefficients.

An interesting opportunity to search for $\alpha$-variation is
associated with M1 transitions between fine-structure (FS) levels of
ground multiplets. These lines are observable for very large
cosmological redshifts, up to $z\sim 10$ \cite{WHD03,MCC05}, and
have high sensitivity to $\alpha$-variation. In the case of
$^2\!P_J$ multiplet of C~\textsc{ii} ion there is only one FS line
(1,0). As a reference one can use microwave molecular lines, which
are insensitive to $\alpha$-variation, but depend on the mass ratio
$\mu$. This way one can put a limit on the variation of the
combination $F=\alpha^2/\mu$ \cite{LRK08}:
 $\Delta F/F=(0.1\pm 1.0)\times 10^{-4}$ at $z=6.42$
(the look back time about 13 Gyr). The signal in this method is
enhanced by the large and different sensitivities of compared lines.
On the other hand, the lines of different species are used, so the
Doppler noise can become a problem when the accuracy is increased.

Many atoms and ions have triplet ground state $^3\!P_J$, and two M1
transitions (1,0) and (2,1) can be observed. The most important
species of this kind for astrophysical observations are Carbon-like
and Oxygen-like ions. In this case one can measure the ratio of
these two frequencies and compare it to the laboratory value. This
way the Doppler noise can be significantly suppressed. However, in
the first approximation within the LS-coupling scheme, the FS
frequencies obey the Land\'{e} rule:
 \begin{align}\label{intr1}
 \omega_{2,1}/\omega_{1,0}=2\,.
 \end{align}
Thus, the frequency ratio does not depend on $\alpha$.

Sensitivity coefficients $\mathcal{Q}$ to $\alpha$-variation are
defined as:
 \begin{align}\label{intr2}
 \Delta\omega/\omega=2\mathcal{Q}\Delta\alpha/\alpha\,.
 \end{align}
In the approximation \eqref{intr1} these coefficients are the same
for both FS transitions:
 \begin{align}\label{intr3}
 \mathcal{Q}_{2,1}=\mathcal{Q}_{1,0}=1\,.
 \end{align}
Equations \eqref{intr1} and \eqref{intr3} break when we take into
account non-diagonal spin-orbit interaction. Equation \eqref{intr1}
also breaks when Breit interaction is taken into account. If the
latter is neglected, one can link experimentally observed violation
of the Land\'{e} rule with the difference in sensitivity
coefficients \cite{KPL08}:
 \begin{align}\label{FS6}
 \Delta \mathcal{Q} =  \mathcal{Q}_{2,1}-\mathcal{Q}_{1,0}
 = \frac{1}{2}\,\left(\frac{\omega_{2,1}}{\omega_{1,0}}\right)-1\,.
 \end{align}
This simple expression predicts significant differences
in sensitivity coefficient of FS transitions, which rapidly grow
with nuclear charge $Z$. In fact, Breit interaction can not be
neglected for the light ions with $Z\lesssim 10$. For this case
\Eref{FS6} significantly overestimates $\Delta \mathcal{Q}$.

For the configuration $ns^2np^2$ one can take Breit interaction into
account within well known semi-empirical theory \cite{Sob79}.
Corresponding results for a number of ions of astrophysical interest
are given in \cite{KPL08}. However, this theory is essentially
one-configurational and does not account for certain correlation
corrections. The latter can be adequately treated only within
relativistic \textit{ab initio} calculations.

In this paper we report \textit{ab initio} calculations of the
sensitivity coefficients $\mathcal{Q}$
for FS transitions in the ground multiplets of Carbon and
Carbon-like Nitrogen, Oxygen, Sodium, Magnesium, and Silicon. Our
results are in a good agreement with analytical estimates
\cite{KPL08}, the differences being typically on the order of 10\%.
All mentioned elements, with exception of Sodium, have several
stable isotopes. Inhomogeneity of isotope distribution and
cosmological evolution of isotope abundances can lead to the isotope
shifts of observed lines in comparison to the laboratory
frequencies. This can cause systematic effects for
$\alpha$-variation studies. Unfortunately, the isotope shifts for
the FS transitions are not known. For this reason we calculated
isotope mass shift coefficients $k_\mathrm{ms}$
for all ions with several isotopes. For the
ions considered here the volume shift is much smaller and can be
neglected.


Among the light elements C, N and O (CNO), it is mainly carbon,
where the cosmic evolution may change isotopic ratios considerably.
The ${}^{12}$C/${}^{13}$C ratio is about 90 in the solar system, but
it can be as low as 6 in the atmospheres of evolved massive red
supergiants like Alpha Ori \cite{HL84}. The enhanced ${}^{13}$C
abundance relative to ${}^{12}$C is caused by the CNO
cycle operating in massive stars. Therefore, one can expect that in
high redshift objects, where only short-lived, massive stars can
have evolved, the ${}^{12}$C/${}^{13}$C ratio can be much below the
solar value and a ${}^{13}$C fraction of up to 20\% is possible in
the early universe. The question of ${}^{12}$C/${}^{13}$C evolution
has also been discussed by \citet{FMG05} for somewhat later phases
where intermediate mass stars convert ${}^{12}$C into ${}^{13}$C in
the Asymptotic Giant Brunch. Here peak values of
${}^{13}$C/${}^{12}$C of 0.06 are predicted. For nitrogen, the
${}^{14}$N/${}^{15}$N ratio is always large so that isotopic shifts
are probably not important. However, in the Murchison meteorite, the
${}^{14}$N/${}^{15}$N ratio varies in SiC grains between $10^2$ and
$4\times 10^3$, and the lower values of  ${}^{14}$N/${}^{15}$N have
been found in grains with low ${}^{12}$C/${}^{13}$C \cite{HAZ94}.
For oxygen, the solar system value ${}^{16}$O/${}^{17}$O is about
2630, but this ratio can be lower in CNO processed material by a
factor of 5 (Alpha Ori, \cite{HL84}). This means that there may be
some cosmic evolution of isotopic ratios in N and O, however at a
much lower level than in C, and not detectable at the present
achievable spectral resolution. For the evolution of the isotopic
ratios of Mg and Si, which have both several abundant isotopes, we
refer to the works of \citet{AMK04} and \citet{FMG05}.


\subsection{Details of the calculations}\label{sec_details}

We used Dirac-Breit Hamiltonian and all-electron configuration
interaction (CI) method. It was shown in Ref. \cite{KPL08} that
deviation from the Land\'{e} rule was caused primarily by
non-diagonal spin-orbit interaction between the levels $^3\!P_{0,2}$
and $^1\!S_0,\,{}^1\!D_2$, while the level $^3\!P_1$ remained
practically unperturbed. Therefore, it was essential that the theory
accurately reproduced the spacings for all 5 levels of the ground
configuration. Below we present our results for all levels of
interest and compare theoretical energies with the experimental data
from NIST \cite{NIST}.

Most calculations were done with Dirac-Sturmian basis set
\cite{TL03,Tup05}, which included orbitals up to $9s$, $9p$, $8d$,
$8f$, and $7g$. Configurational space corresponded to triple and
partly quadruple excitations from $1s$- $2s$- and $2p$-shells. We
also did calculations with a much smaller configurational space
which included singles for $1s$-shell and doubles for $2s$- and
$2p$-shells, but on a somewhat longer basis set. In this smaller
calculation we neglected retardation part of the Breit interaction.
We found out that retardation corrections were very small. Expansion
of the configurational space was more important and led to better
agreement with the experiment for the transition frequencies. On the
other hand, the difference for the sensitivity coefficients was not
dramatic (see below). Correlation effects were most important for
neutral Carbon and less important for the Carbon-like ions.

\begin{table*}[tbh]
\caption{Energies in cm$^{-1}$ and $\mathcal{Q}$-factors for the
levels of the configuration $2s^22p^2$ in respect to the ground
state $^3\!P_0$. Experimental frequencies are taken from
Ref.~\cite{NIST}.} \label{tab_E}
\begin{tabular}{l|ddc|ddc|dcc|dcc}
 \hline\hline
 \multicolumn{1}{c|}{Ion}
 &\multicolumn{3}{c|}{$^3\!P_1$}
 &\multicolumn{3}{c|}{$^3\!P_2$}
 &\multicolumn{3}{c|}{$^1\!D_2$}
 &\multicolumn{3}{c}{$^1\!S_0$}
 \\
 &\multicolumn{1}{c}{Exper.} &\multicolumn{1}{c}{Theor.} &{$\mathcal{Q}$}
 &\multicolumn{1}{c}{Exper.} &\multicolumn{1}{c}{Theor.} &{$\mathcal{Q}$}
 &\multicolumn{1}{c}{Exper.} &\multicolumn{1}{c}{Theor.} &{$\mathcal{Q}$}
 &\multicolumn{1}{c}{Exper.} &\multicolumn{1}{c}{Theor.} &{$\mathcal{Q}$}
 \\
  \hline
C~\textsc{i}   & 16.42  & 16.4    & 0.9998 & 43.41  & 43.3  & 0.9948
               &10192.6 & 10435   & 0.0023 &21648.0 & 21879 & 0.0014   \\[1.5mm]
N~\textsc{ii}  & 48.7   & 48.3    & 1.0086 & 130.8  & 129.3 & 1.0004
               &15316.2 & 15605   & 0.0052 &32688.8 & 33036 & 0.0031   \\[1.5mm]
O~\textsc{iii} & 113.2  &  112.8  & 1.0197 & 306.2  & 304.9 & 1.0040
               &20273.3 &  20535  & 0.0099 &43185.7 & 43540 & 0.0058   \\[1.5mm]
Na~\textsc{vi} & 698    & 693     & 1.0671 & 1859   & 1850  & 1.0132
               &35498   & 35719   & 0.0375 &74414   & 74795 & 0.0215   \\[1.5mm]
Mg~\textsc{vii}& 1107   & 1108    & 1.0891 & 2924   & 2920  & 1.0172
               &40948   & 41192   & 0.0534 &85153   & 85571 & 0.0306   \\[1.5mm]
Si~\textsc{ix} & 2545   & 2532    & 1.1403 & 6414   & 6394  & 1.0254
               &52926   & 53185   & 0.0990 &107799  & 108253& 0.0573   \\
  \hline\hline
\end{tabular}
\end{table*}

Comparison of the results for frequencies and $\mathcal{Q}$-factors
from the calculations with two configurational spaces described
above confirmed that the smaller space was sufficiently saturated.
That allowed us to use it for the calculations of the isotope shifts
and significantly reduce the computational costs.

In order to determine $\mathcal{Q}$-factors we calculated atomic
energy levels for three or five values of $\alpha$ in the vicinity
of its physical value and performed numerical differentiation. Three
point differentiation was used in most previous calculations
\cite{DFW99a,DFK02}. Here we were interested in the small
differences in $\mathcal{Q}$-factors for the FS levels, so we
compared three-point differentiation with more expensive five-point
differentiation. The difference appeared to be rather small.

Calculations of the mass shift coefficients are similar to
calculations of $\mathcal{Q}$-factors \cite{BDF03,BDFK04}. The
Hamiltonian $H_\mathrm{ms}$, which describes mass shift, is added to
atomic Hamiltonian with the coefficient $\lambda$:
$H_\lambda=H+\lambda H_\mathrm{ms}$. After solving eigenvalue
problem for $H_\lambda$ one finds the mass shift coefficient by
numerical differentiation, $k_\mathrm{ms}=\partial
E_\lambda/\partial \lambda$.

For the mass shift we use relativistic theory developed in
\cite{SA94,ASY95,Sha98,Tup03}. For optical transitions in light
atoms the relativistic corrections to mass shift are usually small
\cite{KK07}. However, here we are interested in the infrared FS
transitions. As long as the fine-structure disappears in the
non-relativistic limit, the non-relativistic theory of the isotope
shift is not applicable here.

In the non-relativistic approximation the total mass shift is a sum
of the normal mass shift (NMS) and the specific mass shift (SMS).
The NMS contribution comes from the substitution of the electron
mass with the reduced mass, and respective coefficient is simply
proportional to the transition frequency:
$k_\mathrm{nms}^\mathrm{nr}=\omega/1823$ (a.u.). The SMS
contribution is described by the two-electron operator $\sum_{i\neq
k}\bm{p}_i \cdot \bm{p}_k$ and has to be calculated numerically. In
the relativistic theory the NMS and SMS contributions correspond to
one-electron ($i=k$) and two-electron ($i\neq k$) parts of the
relativistic mass shift operator:
\begin{multline}\label{ms_rel}
 H_\mathrm{MS} = \sum_{i,k}
 \left(\bm{p}
 - \frac{\alpha Z}{2r}
 \left[\bm{\alpha}+
 (\bm{\alpha}\cdot\hat{\bm{r}})\hat{\bm{r}}\right]\right)_i
\\
 \cdot
\left(\bm{p}
 - \frac{\alpha Z}{2r}
 \left[\bm{\alpha}+
 (\bm{\alpha}\cdot\hat{\bm{r}})\hat{\bm{r}}\right]\right)_k,
\end{multline}
which is correct to the second order in $\alpha Z$. Here NMS
contributions have to be calculated numerically on the same footing
as SMS \cite{SA94,ASY95,Sha98,Tup03,KK07}.

\begin{table*}[bth]

\caption{The differences of the sensitivity coefficients $\Delta
\mathcal{Q}$ of the FS emission lines within the ground multiplet
$^3\!P_J$ for the most abundant Carbon-like ions. \textit{Ab initio}
results of this work are compared with semi-empirical results
\cite{KPL08} based on the theory \cite{Sob79}. In addition to the
large CI we also made a much smaller CI as described in the text.}
\label{tab_Q}
\begin{tabular}{ldddddddd}
\hline\hline\\[-7pt]
 \multicolumn{1}{c}{Ion}
 &\multicolumn{2}{c}{Transition $a$ (1,0)}
 &\multicolumn{2}{c}{Transition $b$ (2,1)}
 &\multicolumn{1}{c}{$\omega_b/\omega_a$}
 &\multicolumn{3}{c}{$\Delta \mathcal{Q}=\mathcal{Q}_b-\mathcal{Q}_a$} \\
 &\multicolumn{1}{c}{$\lambda_a$ ($\mu$m)}
 &\multicolumn{1}{c}{$\omega_a$ (cm$^{-1}$)}
 &\multicolumn{1}{c}{$\lambda_b$ ($\mu$m)}
 &\multicolumn{1}{c}{$\omega_b$ (cm$^{-1}$)}
 &&\multicolumn{1}{c}{Ref.\cite{KPL08}}
 &\multicolumn{1}{c}{Large CI}
 &\multicolumn{1}{c}{Small CI} \\
  \hline\\[-5pt]
C~\textsc{i}   & 609.1 & 16.40 & 370.4 &   27.00 & 1.646 & -0.008 & -0.0081 & -0.0090 \\
N~\textsc{ii}  & 205.3 & 48.70 & 121.8 &   82.10 & 1.686 & -0.016 & -0.0132 & -0.0134 \\
O~\textsc{iii} &  88.4 &113.18 &  51.8 &  193.00 & 1.705 & -0.027 & -0.0250 & -0.0247 \\
Na~\textsc{vi} &  14.3 & 698   &   8.6 &  1161   & 1.663 & -0.091 & -0.0861 & -0.0843 \\
Mg~\textsc{vii}&   9.0 & 1107  &   5.5 &  1817   & 1.641 & -0.12  & -0.116  & -0.114  \\
Si~\textsc{ix} &   3.9 &2545.0 &   2.6 & 3869    & 1.520 & -0.21  & -0.190  & -0.188  \\
  \hline\hline
\end{tabular}
\end{table*}

\subsection{Numerical results and discussion}\label{sec_results}

Results of our calculations of the energies and sensitivity
coefficients $\mathcal{Q}$ for the levels of the ground
configuration $1s^2 2s^2 2p^2$ are presented in \tref{tab_E}. All of
them are given in respect to the ground state $^3\!P_0$.
Calculations are done for the ``large'' configurational space
described in \sref{sec_details}. This space includes $\sim 1.2\times
10^5$ relativistic configurations.

We see that for the levels of the ground multiplet the
$\mathcal{Q}$-factors are close to unity, as expected. For two other
levels of the ground configuration the $\mathcal{Q}$-factors are
small. Deviation from unity in the former case and deviation from
zero in the latter are of the same order of magnitude and rapidly
grow with nuclear charge $Z$. Here we are mostly interested in the
differences of the $\mathcal{Q}$-factors for the FS transitions
(1-0) and (2-1). These differences are given in \tref{tab_Q}.

From \tref{tab_Q} one can see that our numerical results are in good
agreement with semi-empirical estimates from Ref.~\cite{KPL08}.
Calculations with ``large'' and ``small'' CI described above are
very close to each other. Except for the neutral Carbon, the
difference between two numerical calculations is much smaller than
their difference from semi-empirical values. Note, that for the
large CI we used five-point differentiation method and three-point
method for the small CI. Therefore, we can conclude that small CI is
already sufficiently saturated and that three-point differentiation
is also sufficiently accurate. Because of that, in the following
calculations of the mass shifts we use this significantly cheaper
method (small CI includes $\sim 10000$ relativistic configurations).

\begin{table}[tbh]
\caption{Coefficients $k_\mathrm{nms}$ and $k_\mathrm{sms}$ for
Carbon-like ions. Mass shift is calculated in respect to the ground
state $^3\!P_0$. All numbers are in
$(\mathrm{GHz}\times\mathrm{amu})$. For NMS we also give
non-relativistic value $k_\mathrm{nms}^\mathrm{nr}=\Delta
E_\mathrm{exper}(\mathrm{a.u.})\times 3609.6$. }
 \label{tab_ms1}
\begin{tabular}{cldddd}
\hline\hline
 &&\multicolumn{1}{c}{$^3\!P_1$}
  &\multicolumn{1}{c}{$^3\!P_2$}
  &\multicolumn{1}{c}{$^1\!D_2$}
  &\multicolumn{1}{c}{$^1\!S_0$} \\
  \hline
\\[-2mm]
 C~\textsc{i}
 & $k_\mathrm{nms}            $& -0.09 & -0.42 & 178.8 & 369.7  \\
 & $k_\mathrm{nms}^\mathrm{nr}$&  0.27 &  0.71 & 167.6 & 356.0  \\
 & $k_\mathrm{sms}            $&  0.51 &  1.75 &-155.2 &-151.2  \\[2mm]
 N~\textsc{ii}
 & $k_\mathrm{nms}            $& -0.31 & -1.26 & 263.5 & 553.2  \\
 & $k_\mathrm{nms}^\mathrm{nr}$&  0.80 &  2.15 & 251.9 & 537.6  \\
 & $k_\mathrm{sms}            $&  1.35 &  4.42 &-179.4 &-167.4  \\[2mm]
 O~\textsc{iii}
 & $k_\mathrm{nms}            $& -0.09 & -1.43 & 343.3 & 725.2  \\
 & $k_\mathrm{nms}^\mathrm{nr}$&  1.86 &  5.04 & 333.4 & 710.2  \\
 & $k_\mathrm{sms}            $&  2.56 &  8.79 &-197.0 &-165.6  \\[2mm]
 Mg~\textsc{vii}
 & $k_\mathrm{nms}            $&  7.33 & 14.51 & 661.9 & 1398.3  \\
 & $k_\mathrm{nms}^\mathrm{nr}$& 18.21 & 48.09 & 673.4 & 1400.4  \\
 & $k_\mathrm{sms}            $& 21.58 & 62.86 &-221.5 &  131.1  \\[2mm]
 Si~\textsc{ix}
 & $k_\mathrm{nms}            $& 19.83 & 43.96 & 834.0 & 1748.3  \\
 & $k_\mathrm{nms}^\mathrm{nr}$& 41.86 &105.49 & 870.4 & 1772.9  \\
 & $k_\mathrm{sms}            $& 45.34 &124.24 &-197.6 &  468.3  \\
  \hline\hline
\end{tabular}
\end{table}

\begin{table}[tbh]
\caption{Frequency shifts $\delta \nu=\nu^{A'}-\nu^A$ for FS
transitions $(J'-J)$ in isotopes $A'$ and $A$ of C-like ions. The
last two columns give respective velocity shifts $\delta V =
-\delta\nu/\nu\times c$, where $c$ is the speed of light.}
 \label{tab_ms2}
\begin{tabular}{cccdddd}
\hline\hline
                &    &    &\multicolumn{2}{c}{$\delta \nu$ (GHz)}
                                                    &\multicolumn{2}{c}{$\delta V$ (km/s)} \\
   Ion          &$A'$&$A $&\multicolumn{1}{c}{$(0-1)$}
                                        &\multicolumn{1}{c}{$(1-2)$}
                                                    &\multicolumn{1}{c}{$(0-1)$}
                                                               &\multicolumn{1}{c}{$(1-2)$}\\
C~\textsc{i}    & 13 & 12 &  -0.00278   &  -0.00602 &    1.691 &   2.228  \\
N~\textsc{ii}   & 15 & 14 &  -0.00498   &  -0.01006 &    1.024 &   1.225  \\
O~\textsc{iii}  & 17 & 16 &  -0.00909   &  -0.01798 &    0.803 &   0.932  \\
                & 18 & 16 &  -0.01717   &  -0.03396 &    1.517 &   1.760  \\
Mg~\textsc{vii} & 25 & 24 &  -0.04819   &  -0.08076 &    0.435 &   0.444  \\
                & 26 & 24 &  -0.09267   &  -0.15531 &    0.837 &   0.855  \\
Si~\textsc{ix}  & 29 & 28 &  -0.08026   &  -0.12689 &    0.315 &   0.328  \\
                & 30 & 28 &  -0.15517   &  -0.24532 &    0.610 &   0.634  \\
  \hline\hline
\end{tabular}
\end{table}

Our results for the mass shifts are summarized in
Tables~\ref{tab_ms1} and~\ref{tab_ms2}. \tref{tab_ms1} gives
coefficients $k_\mathrm{nms}$ and $k_\mathrm{sms}$, while
\tref{tab_ms2} presents full mass shifts for given isotope pairs. We
see that for the FS transitions from the ground state the NMS
coefficients calculated with the help of relativistic operator
\eqref{ms_rel} differ very strongly from the non-relativistic
values. Interestingly, the difference is bigger for the light ions,
where even the sign of the non-relativistic approximation is wrong.
Relativistic NMS coefficient becomes positive between $Z=8$ (Oxygen)
and $Z=9$ (Fluorine), but remains significantly smaller than
non-relativistic predictions. Note that for two optical transitions
between levels of different multiplets the relativistic and
non-relativistic values are very close. This result confirms that
non-relativistic theory of the mass shift works nicely for optical
transitions, but is not applicable to FS transitions.

\tref{tab_ms1} shows that SMS coefficients also significantly change
along the isoelectronic sequence. The NMS coefficients are mostly
sensitive to relativistic effects, which depend on the parameter
$\alpha Z$. The SMS coefficients are more sensitive to correlations.
The latter are governed by the parameter $1/Z$ and decrease along
the sequence. It is clearly seen from \tref{tab_ms2} that full mass
shift for the FS transitions behaves more smoothly, than NMS and SMS
parts. This may mean that in the relativistic theory there is no
good reason to separate mass shift into NMS and SMS parts. The last
two columns of \tref{tab_ms2} give apparent velocity (Doppler)
shifts in astrophysical observations, which correspond to the
frequency shifts from two previous columns. We see that velocity
shifts for two FS transitions significantly differ for light ions,
but become almost equal for heavier ions. For Mg~\textsc{vii} and
Si~\textsc{ix} the isotope shifts practically cancel out from the
frequency ratio $\omega_{2,1}/\omega_{1,0}$. Consequently, for these
ions the isotope shift does not lead to noticeable systematic effect
in $\alpha$-variations studies.

\subsection{Conclusions}\label{conclusions}

FS transitions can be observed in far infrared waveband for very
distant astrophysical objects with redshifts $z\sim 10$. This makes
them very promising candidates for the study of possible
$\alpha$-variation on the cosmological timescale. We performed
\textit{ab initio} calculation of the sensitivity coefficients
$\mathcal{Q}$ to $\alpha$-variation for the FS transitions within
ground multiplet $^3\!P_J$ of Carbon-like light ions. These
calculations confirmed that the differences in sensitivity for FS
transitions $\Delta\mathcal{Q}$ is of the same order as
$\Delta\mathcal{Q}$ for optical transitions in the same ion. In both
cases $\Delta\mathcal{Q}$ rapidly grows with $Z$. In optical
waveband the most dangerous systematic effect is associated with the
isotope shifts. We calculated mass shifts for FS transitions and
found that isotope shift is rather large for light ions
C~\textsc{i}, N~\textsc{ii}, and O~\textsc{iii}, but practically
cancels out for Mg~\textsc{vii} and Si~\textsc{ix}. At the same time
these heavier ions, together with the Na~\textsc{vi} ion, which has
only one stable isotope, have higher sensitivity to
$\alpha$-variation.

It was pointed out in \sref{sec_intro} that in the solar system
light elements C, N, and O have only one dominant isotope. However,
in the early universe the abundances of $^{12}$C and $^{13}$C could
be significantly different. At the same time, the isotope shift for
the FS transitions in C~\textsc{i} is larger than for other C-like
ions, while sensitivity to $\alpha$-variation is lower. We conclude
that the method suggested in \cite{KPL08} may not work for
C~\textsc{i} because of the small sensitivity and large systematic
effects associated with isotope shifts. For C-like ions the
sensitivity to $\alpha$-variation grows with the nuclear charge $Z$,
while isotope shifts rapidly decrease. Moreover, for heavier ions
the isotope shifts almost cancel out for the ratio of the FS
frequencies. Another dangerous systematic effect from the Doppler
noise is significantly suppressed for the pairs of lines of the same
species. Therefore, observations of the FS transitions in heavier
C-like ions can be used as a sensitive tool for the search of
$\alpha$-variation at large redshifts.

\begin{acknowledgments}
This research is partly supported by the DFG projects SFB 676
Teilprojekt C4 and the RFBR grants No. 08-02-00460. MGK gratefully
acknowledges the hospitality of Hamburger Sternwarte.
\end{acknowledgments}


\end{document}